\documentclass[10pt,conference]{IEEEtran}
\newcommand{\ie}{{\it i.e.}}

\newcommand{\argmin}{\mathop{\rm argmin}}

\newcommand{\SCNR}{\mathop{\mathsf{SCNR}}}
\newcommand{\SQNR}{\mathop{\mathsf{SQNR}}}

\usepackage[dvips]{graphicx}
\usepackage[usenames,dvipsnames]{color}
\usepackage{amssymb,amsmath}
\usepackage{multirow}

\begin{document}
\title{On the Capacity of the Discrete-Time Channel with Uniform Output Quantization}
\author{Yiyue Wu \\Electrical Engineering\\ Princeton University,\\ yiyuewu@princeton.edu \and
Linda M. Davis \\Institute of Telecommunication Research\\ University of South Australia\\linda.davis@unisa.edu.au \and
Robert Calderbank \\Electrical Engineering\\ Princeton University,\\ calderbk@math.princeton.edu }
\maketitle
\begin{abstract}
This paper provides new insight into the classical problem of determining both the capacity of the discrete-time channel with uniform output quantization and the capacity achieving input distribution. It builds on earlier work by Gallager and Witsenhausen to provide a detailed analysis of two particular quantization schemes. The first is \emph{saturation} quantization where overflows are mapped to the nearest quantization bin, and the second is \emph{wrapping} quantization where overflows are mapped to the nearest quantization bin after reduction by some modulus. Both the capacity of wrapping quantization and the capacity achieving input distribution are determined. When the additive noise is gaussian and relatively small, the capacity of saturation quantization is shown to be bounded below by that of wrapping quantization. In the limit of arbitrarily many uniform quantization levels, it is shown that the difference between the upper and lower bounds on capacity given by Ihara is only 0.26 bits.
\end{abstract}
\section{Introduction}
Modern communication systems rely on digital processing of data where the received signals are quantized by analog to digital converters (ADC). In this paper, we focus on quantization with a finite number of output levels which is referred to as finite-level quantization. We also consider the limiting case where the number of the quantization output levels goes to infinity and we refer to it as infinite-level quantization.

For finite-level quantization, Witsenhausen \cite{Witsenhausen} used Dubins' theorem \cite{Dubins} to show that the capacity of a discrete-time memoryless channel with output cardinality $N$, under a peak power constraint is achievable by a discrete input with at most $N$ nonzero probability mass points. The authors in \cite{Singh} considered average power constraint instead and showed that the capacity is achievable by a discrete input with at most $N+1$ nonzero probability mass points. We note that Gallager first showed that the number of input nonzero probability mass points need not exceed the number of quantization output levels \cite{Gallager} (p. 96, Corollary 3). However, the optimal input distribution and channel capacity for this system remain open.

In order to better understand the channel with finite-level quantization, this paper compares \emph{saturation} quantization and \emph{wrapping} quantization which deal with ADC input overflow differently. When an overflow occurs, a saturation quantizer simply maps it to its nearest quantization bin while a wrapping quantizer maps it to its nearest quantization bin after a modulo operation (see Section \ref{MQsystems}). This paper derives the exact capacity and capacity achieving input distribution for a discrete-time channel with uniform wrapping quantization regardless of the noise distribution. This paper further shows that when the additive noise is gaussian and relatively small, the capacity of uniform wrapping quantization is a lower bound for the capacity of uniform saturation quantization. We also provide an input distribution which approaches the capacity of saturation quantization in the limit when the number of quantization levels is relatively large.

We also analyze the capacity of discrete-time channel under uniform infinite-level quantization. We observe that the capacity can be nicely approximated by the lower bound and upper bound derived by Ihara \cite{Ihara}.

The rest of this paper is organized as follows. Section \ref{SystemModel} introduces the system model and different types of quantizers. Section \ref{CapacityFinite} analyzes the system capacity under uniform finite-level output quantization. Section \ref{CapacityInfinite} discusses the system capacity under uniform infinite-level output quantization. Conclusions are provided in Section \ref{conclusion}.

\section{System Model}
\label{SystemModel}
\subsection{Receive Structure}
We consider a real single-input single-output discrete-time system. The received signal after quantization at the receiver is
\begin{equation}
\label{modelSISO} \hat{y}=\textrm{Q}(hx+n)
\end{equation}
where $x$ is the transmitted signal, $h$ is the channel gain known at the receiver, $n$ is the additive noise, $h$ and $n$ are modeled as certain distributions with zero means and variances, $\sigma_h^2$ and $\sigma_n^2$ respectively and $\textrm{Q}(\cdot)$ is the quantization operation.

\subsection{Quantization}
We first consider quantization with a finite number of output levels and we further consider the extreme case when the number of quantization output levels goes to infinity. For the sake of simplicity,  we refer to these as finite-level quantization and infinite-level quantization respectively.

\subsubsection{Finite-level Quantization}
\label{MQsystems}
For this type of quantization, the output alphabet $\mathcal{O}$ is finite with cardinality $N$:
\begin{equation}
\mathcal{O}=(Y_1,Y_2\cdots,Y_N).
\end{equation}
One approach to finite-level quantization is to map the received signal $y=hx+n$ to one point on the alphabet $\mathcal{O}$ by modulo
and rounding operation as
\begin{equation}
\label{wrapping}
\displaystyle
\hat{y}=\argmin_{Y_i\in \mathcal{O}}\min_{k\in \mathbb{Z}}|kT+Y_{i}-(hx+n)|
\end{equation}
where $T$ is the modulo period. This type of quantization corresponds to the case that once data overflow occurs,  the quantizer keeps its $\log{N}$ least significant bits and ignores the overflow bits. It is referred to as \emph{wrapping} in the Matlab fixed-point toolbox.

For a finite output alphabet, a more common method of handling overflow, \emph{saturation}, has the following operation
\begin{equation}
\displaystyle
\label{saturate}
\hat{y}=\argmin_{Y_i\in \mathcal{O}}|Y_{i}-(hx+n)|.
\end{equation}
In this case, when the overflow occurs, the received signal $y=hx+n$ is mapped to the largest (or smallest) point on the alphabet $\mathcal{O}$.

\textbf{Remark:} Tomlinson filtering \cite{Tomlinson} employs modular arithmetic to enable symbol by symbol decoding in partial response signaling. The associated increase in transmitted signal power (see \cite{Mazo}) is the counterpart of quantization error in this paper.

\subsubsection{Infinite-level Quantization}
We consider that the continuous output ($y=hx+n$) is rounded by quantization and the output alphabet is infinitely countable. In this case,
the quantized received signal $\hat{y}$ can be expressed as
\begin{equation}
\label{Qerror}
\hat{y}=y+\delta y
\end{equation}
where $\delta y$ is the corresponding output quantization error.

Using equation (\ref{Qerror}), equation (\ref{modelSISO}) can be written as
\begin{equation}
\label{model1} \hat{y}=hx + n + z
\end{equation}
where $z=-\delta y $. We consider that $z$ has zero
 mean and variance $\sigma_z^2$ and is independent of channel noise. We define the signal to channel noise ratio as
\begin{equation}
\label{SCNR}
\SCNR=\frac{ E_x\sigma_h^2}{\sigma_n^2}
\end{equation}
and the signal to quantization noise ratio as
\begin{equation}
\label{SQNR}
\SQNR=\frac{ E_x\sigma_h^2}{\sigma_z^2}.
\end{equation}
where $E_x$ is the average constellation power.

\section{Capacity of Systems with Finite-level Uniform Output Quantization}
\label{CapacityFinite}
Without loss of generality, we consider the channel gain $h=1$ in this section.  We study two types of uniform finite-level quantization (wrapping and saturation) with $N$ outputs $Y_i$ given by \[Y_i=(i-1)p,\;\;\;i=1,\cdots,N\] where $p$ is the quantization resolution. For wrapping quantization, the modulo period is $T=Np$ in equation (\ref{wrapping}).

The system capacity can be calculated as
\begin{equation}
\label{}
C=\max_{\textrm{P}(x)}\; \mathcal{I}(x;\hat{y})
\end{equation}
where \(\mathcal{I}(x;\hat{y})=\mathcal{H}(\hat{y})-\mathcal{H}(\hat{y}|x)\) and \(\mathcal{H}(v)\) represents the entropy of variable $v$.

We note that the capacity associated with saturation quantization remains as an open question \cite{Singh}, and neither the capacity nor the capacity achieving input is well understood. However, for wrapping quantization, we are able to derive both the capacity and capacity achieving input distribution. This new result provides an approximation to the capacity of saturation quantization since we are able to show that this quantity is bounded below by the capacity of wrapping quantization when the additive noise is gaussian and relatively small.

\subsection{Review on Related Works}
Wrapping quantization has been understudied compared to general finite-level quantization such as \cite{Witsenhausen,Singh}. In \cite{Witsenhausen}, Witsenhausen considered a stationary discrete-time memoryless channel with a continuous input subject to peak power constraint and a discrete output $\hat{y}\in \{Y_1,Y_2\cdots,Y_N\}$ of finite cardinality $N$ and he proved that if channel transition probability functions $f_i(X) = \Pr(\hat{y}=Y_i|x=X)$ are continuous, then the capacity is achievable by a discrete input distribution with at most $N$ nonzero probability mass points. In \cite{Singh}, the authors considered the same system model but with an average power constraint and they showed that the capacity is achievable by a discrete input with at most $N+1$ nonzero probability mass points. However, neither of these papers characterize the capacity achieving input distribution. In fact, the optimal input and the capacity for this system remain as open questions.  In contrast, for the case of uniform wrapping quantization, both capacity and the optimal input can be established. From Witsenhausen \cite{Witsenhausen}, we can derive lemma 1 easily for the case of wrapping quantization.

\textbf{Lemma 1:} Consider a stationary discrete-time wrapping quantization channel with a continuous input $x$ and a discrete output $\hat{y}\in \{Y_1,Y_2\cdots,Y_N\}$ of finite cardinality $N$. If the the channel transition probability functions are continuous, then the capacity is achievable by a discrete input distribution with at most $N$ mass points.

\textbf{Proof:} Lemma 1 is a direct result of the fact that due to wrapping operation, the input can be treated as confined on the interval $[0, T]$ where $T$ is the underlying modulus.\hfill $\square$

\subsection{Capacity with Uniform Wrapping Quantization}
We now focus on uniform wrapping quantization system with $N$ output and show that the capacity can be achieved by an explicit uniform input distribution with exactly $N$ mass points.

\textbf{Proposition 1:} For a stationary discrete-time channel with wrapping quantization with $N$ finite output as \(Y_i=(i-1)p,\;\;\;i=1,\cdots,N\) and quantizing operation as \[\displaystyle Q(v)=\argmin_{Y_i} \min_{k\in \mathbb{Z}}{|kNp+Y_i-v|},\] the capacity can be achieved by an equiprobable input on $N$ mass points.\\

\textbf{Proof:} Let $u_0\in[0,Np]$ and $\displaystyle g(w)=\mathcal{H}(\hat{y}|x=w)$ such that  $$\displaystyle g(u_0)=\min_{w\in[0,Np]} g(w).$$ Due to modulo operation, we can conclude that
$$\displaystyle g(u_0) = \min_{w\in(-\infty, \infty)} g(w).$$
Since the finite output is in an arithmetic series, there exist another $N-1$ points $$u_i=(u_0+ip)\mod Np,\;\; i=1,\cdots,N-1$$ such that \[g(u_i)=g(u_0).\]

Now we consider the input alphabet as $[u_0,\cdots,u_{N-1}]$, then $\mathcal{H}(\hat{y}|x)$ is minimized. By assigning equiprobability on each mass point, $\mathcal{H}(\hat{y})$ is maximized and $\max \mathcal{H}(\hat{y})=\log{N}$. Therefore, the system capacity is
$$C=\log{N}-\mathcal{H}(\hat{y}|x^*)$$
where $x^*$ is uniformly distributed on the alphabet $[u_0,\cdots,u_{N-1}]$.\hfill $\square$

Proposition 1 shows that for uniform wrapping quantization systems with finite output of cardinality $N$, there exists a uniformly distributed input on $N$ mass points which achieves capacity regardless of the noise distribution. These $N$ mass points may, however, be different for different noise models.

\begin{figure}[h!]
\begin{center}
\resizebox{6cm}{!}{\includegraphics[scale=4.5]{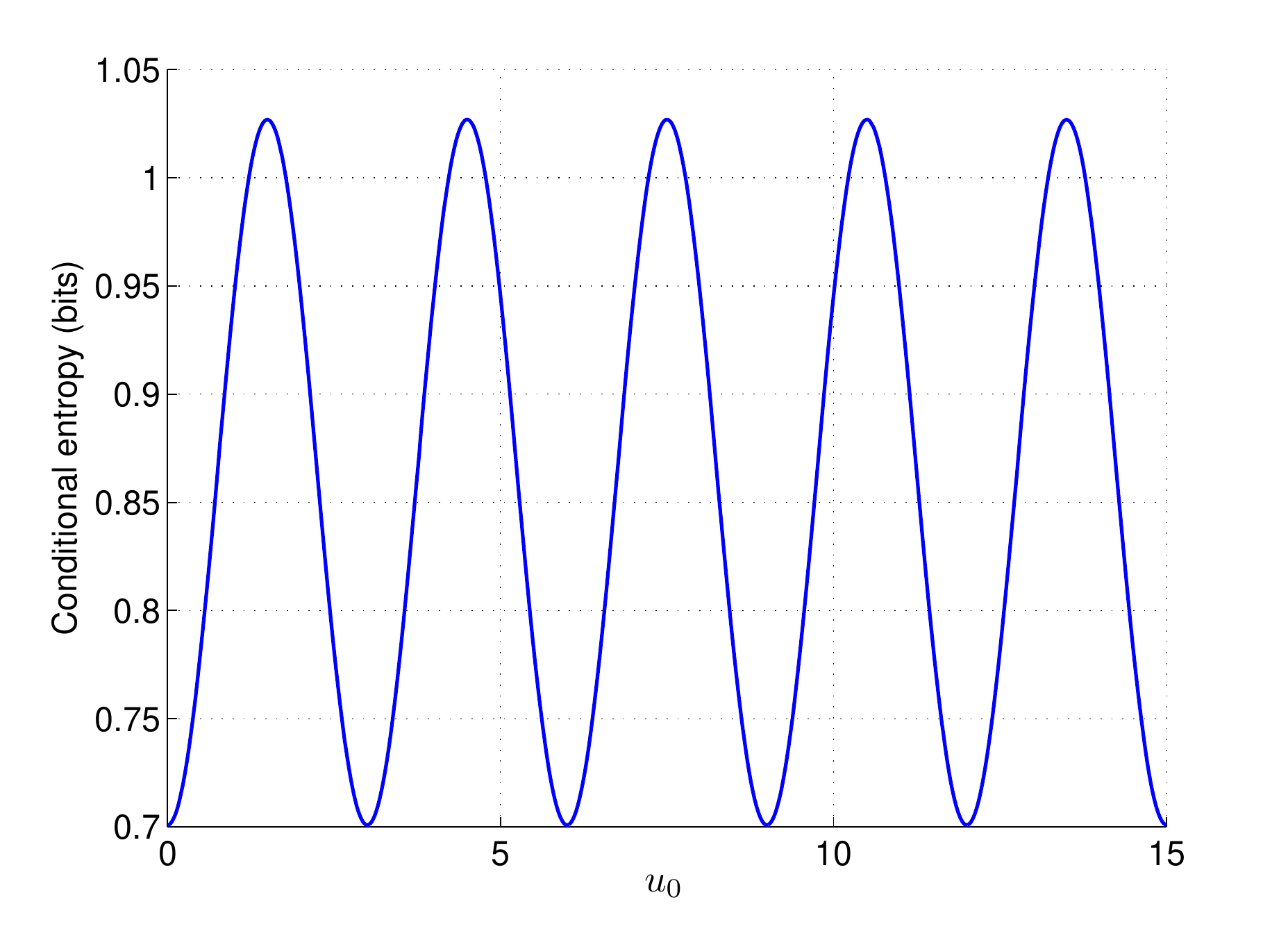}}
\caption{Conditional entropy with standard gaussian noise, $N=5$ and $p=3$.} \label{fig:conditionalentropy}
\end{center}
\end{figure}

The conditional entropy $\mathcal{H}(\hat{y}|x=u_0)$ is illustrated in Fig. \ref{fig:conditionalentropy} when we consider the channel noise $n$ to be gaussian distributed with zero mean and unit variance. In Fig. \ref{fig:conditionalentropy}, the number of output levels is $N=5$ and the quantization resolution is $p=3$. It shows that the conditional entropy is minimized when $u_0$ is at the mass points of the output.

\begin{figure}[h!]
\begin{center}
\resizebox{6cm}{!}{\includegraphics[scale=4.5]{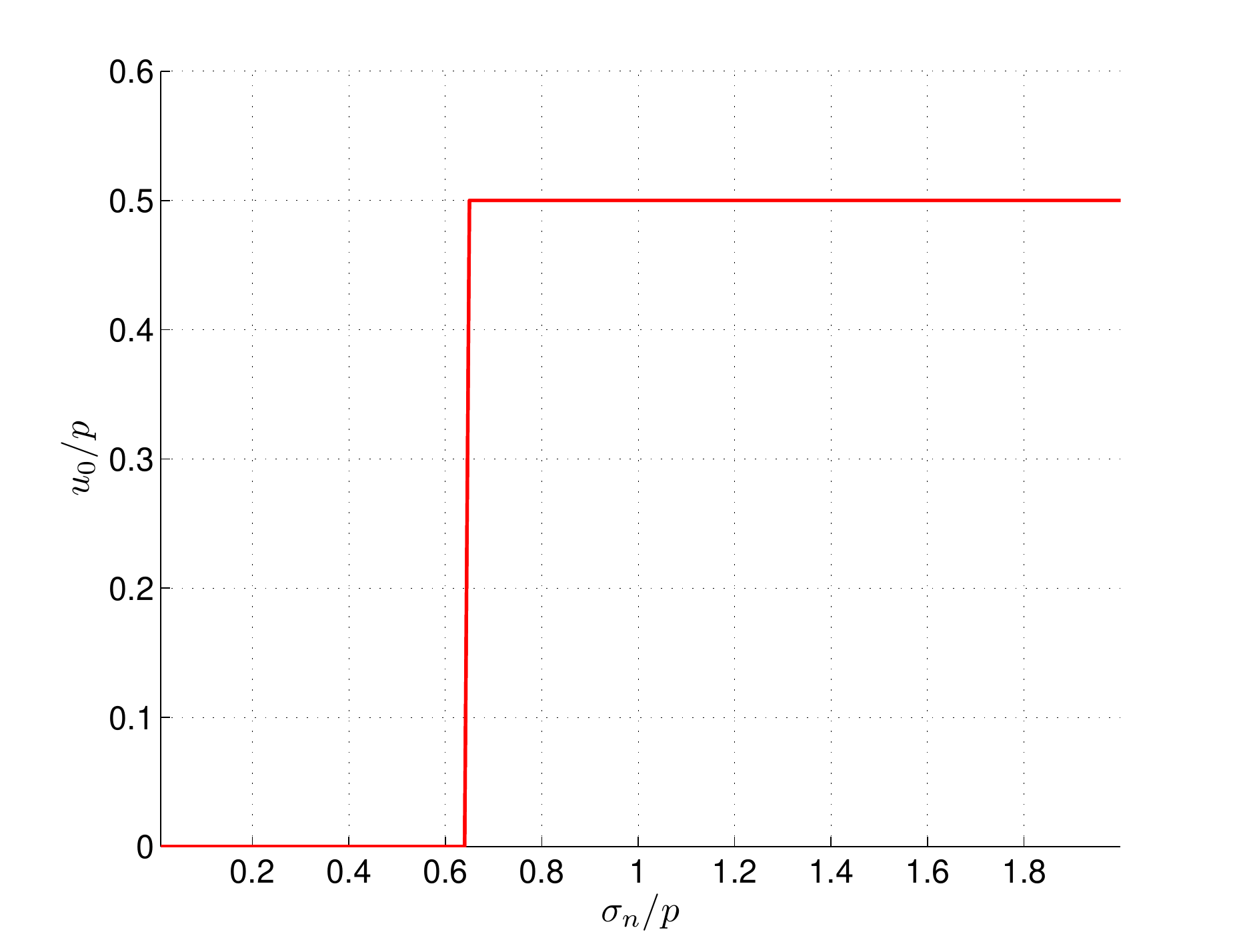}}
\caption{Distribution of $u_0$ with respect to standard deviation of the zero mean gaussian noise.} \label{fig:conditionalentropyminimizer}
\end{center}
\end{figure}

Yet for gaussian noise with relatively large variance and also for other noise distributions, the optimal input might possibly be different. Due to periodicity, we can restrict $u_0$ to be on the interval $[0,p]$. Simulation in Fig. \ref{fig:conditionalentropyminimizer} shows that $u_0$ changes as a step function with respect to the standard deviation of the zero mean gaussian noise (both $u_0$ and the noise standard deviation are normalized with respect to $p$):
\begin{equation}
\label{GnoiseStep}
u_0/p=\left\{\begin{array}{ll}
0 & \sigma_n/p \leq \tau\vspace{2mm}\\
0.5 & \sigma_n/p > \tau.
\end{array} \right.
\end{equation}
where $\tau\approx 0.64$.

For uniformly distributed noise, we have the following result.

\textbf{Proposition 2:} Suppose $n$ is uniformly distributed on $[-\alpha \frac{p}{2}, \alpha \frac{p}{2}]$. For $\alpha \leq 1$, input with equiprobability at output mass points achieves the capacity at $\log{N}$; for $1<\alpha\leq N$, the optimal input alphabet is $$\left\{[0, p, \cdots, (N-1)p]+ \alpha \frac{p}{2} \right\}\mod Np$$ or $$\left\{[0, p, \cdots, (N-1)p]- \alpha \frac{p}{2} \right\}\mod Np$$ which achieve the capacity at $$C=\log{N}-\mathcal{H}(v)$$ where $v$ has mass probability distribution at $\lfloor\alpha\rfloor$ mass points with probability as $$\left(\frac{1}{\alpha}, \cdots, \frac{1}{\alpha}, \frac{\alpha-\lfloor\alpha\rfloor}{\alpha}\right)$$ with $\lfloor\alpha\rfloor$ representing the largest integer smaller than $\alpha$.\\

\textbf{Proof:} When $\alpha \leq 1$, the result is evident.

When $1<\alpha\leq N$, the noise covers the whole decision range for $\lfloor\alpha\rfloor-1$ output mass points and these outputs occur with probability $\frac{1}{\alpha}$ for each. Another two output mass points share the remaining $\frac{\alpha-\lfloor\alpha\rfloor}{\alpha}$ probability. Due to the concavity of the entropy function, $\mathcal{H}(\hat{y}|x)$ is minimized if and only if one mass point of these two takes this remaining probability. In this case, $\mathcal{H}(\hat{y}|x)=\mathcal{H}(v)$ where $v$ has mass
probability distribution at $\lfloor\alpha\rfloor$ mass points with probability as
$$\left(\frac{1}{\alpha}, \cdots, \frac{1}{\alpha}, \frac{\alpha-\lfloor\alpha\rfloor}{\alpha}\right)$$ and the input alphabet is $$\left\{[0, p, \cdots, (N-1)p]+ \alpha \frac{p}{2} \right\}\mod Np$$ or $$\left\{[0, p, \cdots, (N-1)p]- \alpha \frac{p}{2}\right \}\mod Np.$$ With equiprobable input from this alphabet, the capacity is achieved at $C=\log{N}-\mathcal{H}(v)$.

Note, for $\alpha>N$, we can likewise determine the capacity and input distribution, however the effect of the the modulo operation needs to be more carefully considered. \hfill$\square$

%\textbf{Remark:} We can draw the conclusion that for systems with uniform wrapping quantization with finite output cardinality $N$, the capacity can be obtained by a uniform finite input with cardinality $N$. Both the capacity and the optimal input distribution can be determined.

\subsection{Relationship to Capacity with Uniform Saturation Quantization}
For systems with uniform finite-level quantization, we define that the gaussian noise is \emph{weak} if its variance satisfies:
\begin{enumerate}
\label{weakcon}
\item $\sigma_n/p \leq \tau$ where $\tau$ is defined in equation (\ref{GnoiseStep}), and
\item $\Pr(|n|\geq \frac{3p}{2})\approx 0$.
\end{enumerate}

\textbf{Proposition 3:} Consider a gaussian noise corrupted system with input confined on the interval $[0,Np]$ and output from the finite alphabet $ \{Y_1,Y_2\cdots,Y_N|Y_i=(i-1)p\}$. Let $C_s$ be the system capacity under saturation quantization and $C_w$ be the system capacity under wrapping quantization with modulo period $T=Np$. Then if the noise $n$ satisfies the \emph{weak} conditions, we have $$C_s > C_w.$$

\textbf{Proof:} Without loss of generality, we consider the noise has zero mean. Under the weak gaussian noise condition, it is known from the previous subsection that the optimal input distribution for uniform wrapping quantization is uniform distribution on output alphabet and the capacity is \[C_w=\log{N}-\mathcal{H}(v)\] where $v$ has mass probability distribution over $3$ mass points with probability as \[\left(\Pr(|n| \leq p/2),\frac{1}{2}\Pr(|n| > p/2), \frac{1}{2}\Pr(|n| > p/2)\right).\]

Now, we consider applying the same input distribution to saturation quantization. The mutual information is
\begin{equation}
\label{Is}
\mathcal{I}_s(x;\hat{y})=\log{N}-\frac{N-2}{N}\mathcal{H}(v)-\frac{2}{N}\mathcal{H}(u)
\end{equation}
where $u$ is binary distributed with probability as \[\left(\Pr(|n| \leq p/2)+\frac{1}{2}\Pr(|n| > p/2), \frac{1}{2}\Pr(|n| > p/2)\right).\]

Due to concavity of the entropy function, we have $\mathcal{H}(u) < \mathcal{H}(v)$. Therefore, \[C_s\geq \mathcal{I}_s(x;\hat{y}) > C_w.\;\; \]\hfill $\square$

\begin{figure}[h!]
\begin{center}
\resizebox{6cm}{!}{\includegraphics[scale=4.5]{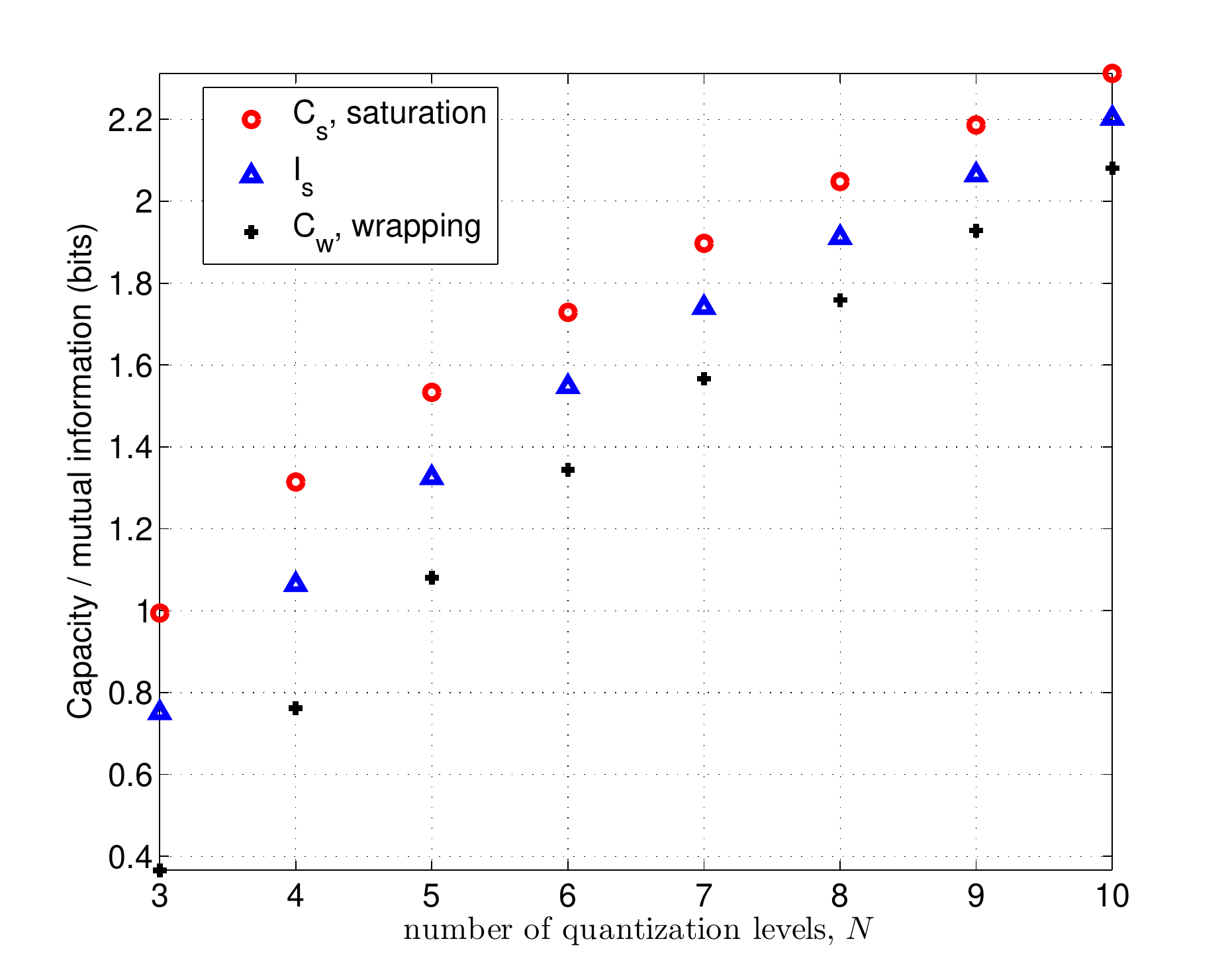}}
\caption{Capacity comparison under weak gaussian noise condition.} \label{fig:capacitycomparison}
\end{center}
\end{figure}

Fig. \ref{fig:capacitycomparison} illustrates the capacities of saturation quantization and wrapping quantization with respect to the number of quantization levels where noise variance $\sigma^2=1$ and quantization resolution $p=2$. The capacity of saturation quantization is obtained by exhaustive search and the mutual information, $\mathcal{I}_s$, is defined in equation (\ref{Is}).

It is observed from Fig. \ref{fig:capacitycomparison} that the gap between capacities of saturation quantization and wrapping quantization decreases as the number of quantization levels increases. And we also observe that $\mathcal{I}_s$ serves as a nice approximation to the capacity of saturation quantization, especially when the number of quantization levels increases.

\section{Capacity of Systems with Infinite-level Uniform Output Quantization}
\label{CapacityInfinite}
Infinite-level uniform quantization models the extreme case for both saturation quantization and wrapping quantization, when the number of quantization levels goes to infinity.
\subsection{Uniform Approximation of Quantization Error}
Infinite-level uniform quantization with step size $p$ (quantization resolution) is a nonlinear process converting continuous signals into discrete signals in a staircase-type relation (see \cite{Proakis}).

In equation (\ref{Qerror}), we assume the received signal $y$ before quantization to be with probability density function $f_y$ and characteristic function $$\phi_y(s)=\mathbb{E}_y[\exp(isy)].$$

The probability density function of $\delta y$ is given by Sripad and Snyder \cite{Sripad}
\begin{equation}
\label{pdfQ}
\small
f_{\delta y} (e) = \left \{ \begin{array}{ll}
\displaystyle  \small \frac{1}{p} + \frac{1}{p} \sum_{k\neq0} \phi_y \left(\frac{2\pi k}{p}\right) \exp\left(\frac{-i2\pi ke}{p}\right),\vspace{2mm} & \;\\
\; & \hspace{-15mm} \small -p/2 \leq e \leq p/2\\
\; & \; \\
0 & \hspace{-15mm} \textrm{otherwise.}
\end{array} \right.
\end{equation}

We have Lemma 2 proved in \cite{yiyueSP}.

\textbf{Lemma 2} \cite{yiyueSP}: If $f_y$ is symmetric with $y=0$ and $\exists\; \alpha>1$ such that $\displaystyle\lim_{u\rightarrow\infty} \phi_y(u)u^{\alpha}=0$, then the distribution
$f_{\delta y}$ of the quantization error $\delta y$ converges to the uniform distribution as the quantization step size goes to zero, \ie
\begin{displaymath}
\lim_{\Delta\rightarrow0} f_{\delta y} (e) = \left \{ \begin{array}{ll}
\frac{1}{p},& -p/2 \leq e \leq p/2\\
\; & \; \\
0 & \textrm{otherwise.}
\end{array} \right.
\end{displaymath}

Based on Lemma 2, we can easily derive proposition 4.

\textbf{Proposition 4:} Consider the system model in equation (\ref{model1}) with $h$ and $n$ as independently gaussian distributed with zero means and variances $\sigma_h^2$, $\sigma_n^2$ respectively. Then $\delta y$ converges to the uniform distribution as the quantization step size goes to zero.

\textbf{Proof:}
Let $g=hx$ and we have
$$\phi_{g}(u)=\mathbb{E}_x\left[\mathbb{E}_h[\exp{(iuhx)}|x]\right] = \mathbb{E}_x\left[\exp(-\frac{\sigma_h^2u^2x^2}{2})\right]\leq 1,$$
so we have $\phi_{y}(u)=\phi_{g}(u)\phi_{n}(u)\leq \phi_{n}(u).$\vspace{2mm}

Since $\phi_{n}(u)=\exp{(-\frac{\sigma_n^2u^2}{2})}$, therefore $\exists\;\; \alpha=2$, such that
$$\lim_{u\rightarrow\infty} \phi_{y}(u)u^{\alpha}=0.$$

We also note that $f_y(a)=f_y(-a)$. Therefore, the conclusion is justified. \hfill $\square$

\subsection{Bounds of Capacity}
We now assume the quantization error $z$ in equation (\ref{model1}) is uniformly distributed with zero mean and variance $\sigma_z^2$, and independent of channel noise $n$.
Then an inequality for the capacity of the quantized system $C$ is given by Ihara \cite{Ihara}
\begin{equation}
\label{capacityInequality} C_0\leq C \leq C_0 + D(P(0,\sigma_z^2 + \sigma_n^2)||\mathcal{N}(0,\sigma_z^2+\sigma_n^2))
\end{equation}
where
\begin{displaymath}
\label{capacity1} C_0= \frac{1}{2}\log \left
(1+\frac{E_x \sigma_{h}^2 }{\sigma_n^2+\sigma_z^2}\right ),
\end{displaymath}
$P(0,\sigma_z^2 + \sigma_n^2)$ represents the distribution of the sum noise ($z+n$), $\mathcal{N}(0,\sigma_z^2+\sigma_n^2)$ represents the gaussian distribution with zero mean, variance $\sigma_z^2+\sigma_n^2$ and $D(P||Q))$ is the divergence defined as
\begin{equation}
\label{divergence} D(P||Q))=\int_{P}\log{\frac{P}{Q}}.
\end{equation}

Note that $C=C_0$ if and only if $z$ follows complex gaussian as $G(0,\sigma_z^2)$ and it is independent of $n$.

The probability density function for the sum noise ($z+n$) can be explicitly written as
$$f_{z+n}(t) = \frac{\Phi(\frac{t+\sqrt{3}\sigma_z}{\sigma_\textbf{n}})-\Phi(\frac{t-\sqrt{3}\sigma_z}{\sigma_\textbf{n}})}{2\sqrt{3}\sigma_z}$$
where $\Phi(\cdot)$ is the cumulative distribution function of standard gaussian distribution.

\begin{figure}[h!]
\begin{center}
\resizebox{6cm}{!}{\includegraphics[scale=4.5]{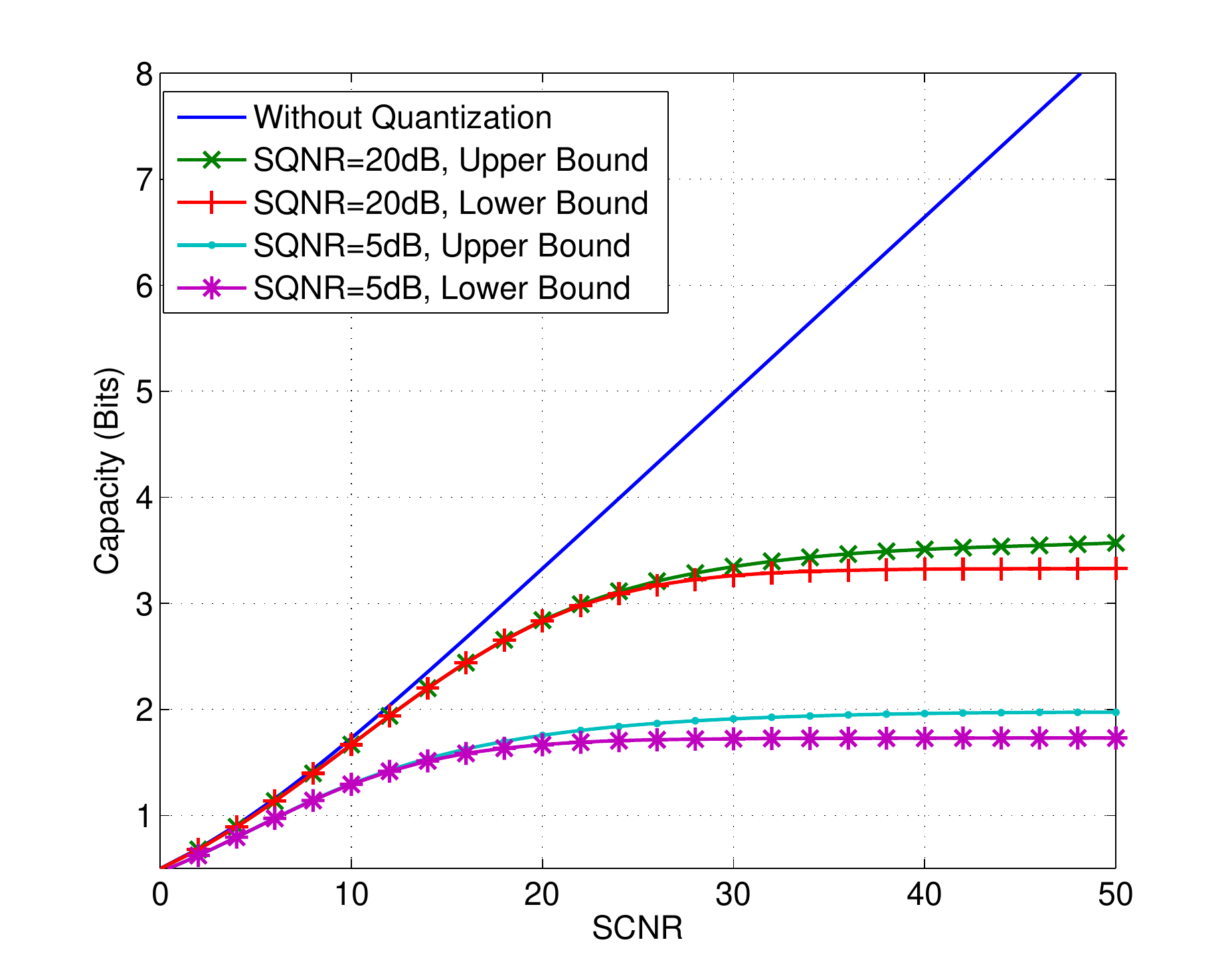}}
\caption{The lower and upper bounds of channel capacity with $\SQNR=5,\; 20$ dB.} \label{fig:capacity}
\end{center}
\end{figure}

Now, we consider the system model in equation (\ref{model1}) with signal to quantization noise ratio fixed as $\SQNR=5,\; 20$ dB and vary the signal to channel noise ratio $\SCNR$. Fig. \ref{fig:capacity} illustrates the lower and upper bounds of channel capacity compared with the channel capacity without quantization. The maximum gap between the lower and upper bounds is
\begin{equation}
\begin{array}{ll}
\max(D(P(0,\sigma_z^2 + \sigma_n^2)||N(0,\sigma_z^2+\sigma_n^2)))\vspace{3mm}\\
 \hspace{35mm}= D(U(0,\sigma_z^2 + \sigma_n^2)||N(0,\sigma_z^2+\sigma_n^2))\vspace{3mm}\\
%  \hspace{35mm}=\frac{1}{2\sqrt{3}} \int_{\sqrt{3}}^{\sqrt{3}} \log_2{\frac{\pi}{\sqrt{3}} \exp{(x^2)}}dx \vspace{3mm}\\
    \hspace{35mm} \approx 0.26\;\; \textrm{bits}
\end{array}
\end{equation}
which is independent of $\SQNR$.

Therefore, either the lower or upper bound is a nice approximation of channel capacity with quantization under the uniform approximation of quantization noise.

\section{Conclusion}
\label{conclusion}
We have investigated the capacity of the discrete-time channel with both finite-level uniform quantization and infinite-level uniform quantization. For finite-level uniform quantization, we have derived the capacity and capacity achieving input distribution associated with uniform wrapping quantization. We have also studied the relationship between the capacities associated with saturation quantization and wrapping quantization. For infinite-level uniform quantization, we have analyzed the quantization error distribution and studied the capacity using the lower and upper bounds by Ihara \cite{Ihara}.

\section{Acknowledgement}
\label{acknowledgement} The authors would like to thank Albert Guillen i Fabregas and Alex Grant
for many helpful suggestions.

% that's all folks
\end{document}